\documentclass[twocolumn,amsmath,amssymb,prl,a4paper]{revtex4}
\usepackage{graphicx}
\usepackage{float}
\usepackage{epsfig}
\usepackage{subfigure}
\usepackage{pstricks}
\usepackage{color}

\newcommand{\Tr}{\mathop\mathrm{Tr}}

\newcommand{\ket}[1]{\left|#1\right>}
\newcommand{\bra}[1]{\left<#1\right|}
\newcommand{\braket}[2]{\left<#1\left|#2\right.\right>}
\newcommand{\expc}[1]{\left<#1\right>}

\date{\today}
\begin{document}
\title{Violation of smooth observable macroscopic realism in a harmonic oscillator}
\author{Amir Leshem and Omri Gat}
\affiliation{Racah Institute  of  Physics, Hebrew University
of Jerusalem, Jerusalem 91904, Israel}
\begin{abstract}
We study the emergence of {macrorealism} in a harmonic oscillator subject to consecutive measurements of a squeezed action. 
We demonstrate a breakdown of {dynamical realism} in a wide parameter range that is maximized in a scaling limit of extreme squeezing, where it is based on measurements of {smooth} observables, implying that macroscopic realism is not valid in the harmonic oscillator. We propose an indirect experimental test of these predictions with entangled photons by demonstrating that local realism in a composite system implies {dynamical realism} in a subsystem.
\end{abstract}
\maketitle
\paragraph{Introduction}
In spite of  prolonged efforts, the question of what constitutes a macroscopic object is still not settled. From the operational point of view the answer is clear: Macroscopic objects are those that are well-described by classical physics. However, it is also widely believed that macroscopic objects are in principle governed by the laws of quantum mechanics, and that classical physics is an emergent phenomenon at an appropriate limit. A common view holds that a macroscopic object becomes classical when it interacts with an environment, that might as well consist of its internal degrees of freedom \cite{zurek}. As a result of decoherence quantum uncertainty is translated into classical ignorance that can be expressed in terms of classical hidden variables. A different approach hangs the emergence of macroscopic realism of closed systems on the impossibility of distinguishing individual quantum levels \cite{kofler1}.

Classical mechanics postulates more than the existence of definite values for all the observables. Since the motion is deterministic, classical observables follow definite trajectories that constrain the different-time joint probability distribution of classical observables. In other words, a hidden variable underpinning of an evolving quantum systems becomes a hidden trajectory dynamical model. If one adds to the hidden trajectory model the assumption (valid in classical mechanics) that a measurement can be performed with an arbitrarily weak perturbation of the system, it is possible to derive Leggett-Garg inequalities (LGIs)  that constrain linear combinations of temporal correlation functions $C(t,t')=\expc{S(t)S(t')}$ of a bounded observable $S$ \cite{lg}. For example, the analog of the Clauser-Horne-Shimony-Holt (CHSH) \cite{chsh} inequality is the LGI $|\expc{LG}|=|C(t_1,t_2)+C(t_2,t_3)+C(t_2,t_3)-C(t_1,t_4)|\le2$. In the formulation of this LGI it is usually assumed that the observable $S$ can assume only the values $\pm1$, but as with the CHSH inequalities \cite{cs}, it is only necessary to assume that $|S|\le1$ to derive the LGI \cite{leggett}. 

Inasmuch as they apply to microscopic states, the assumptions leading to the LGI, that we will summarize by the term \emph{dynamical realism} (DR), are clearly false, and indeed the LGI can be violated for any closed quantum system \cite{kofler2}.  Furthermore, most quantum systems do not allow for a hidden trajectory underpinning, and there breakdown of DR can be demonstrated with a \emph{single} measurement as shown in \cite{nlo} for nonlinear oscillators.
On the other hand, there are quantum systems that admit an \emph{exact} hidden trajectory model, the simplest of which is a harmonic oscillator prepared in a positive Wigner function state, and there the breakdown of DR can be demonstrated with the successive measurement scheme of Leggett and Garg.

The original motivation for the introduction of the LG inequality was to test macroscopic realism (MR). Breakdown of MR follows immediately from LGI violation if $S$ is a dichotomous observable, with support on two macroscopically distinct states undergoing von Neumann measurements. This case was studied previously in \cite{lg,kofler1,kofler2,zela}, but such two-state observables must resolve individual eigenstates, and are therefore singular in the classical limit. In contrast, the harmonic oscillator is a system with a well-behaved classical limit, and it is natural to test MR there with {smooth} observables, like position or momentum, that have a good classical limits as well.
When $S$ is a bounded smooth observable the relative difference in its values for eigenstates that are not macroscopically distinct is by definition small. Therefore, the breakdown of MR is implied by a \emph{large} violation of the LGI, that does not tend to zero in the classical limit.

\begin{figure}[htb]
\subfigure[]{\epsfig{file=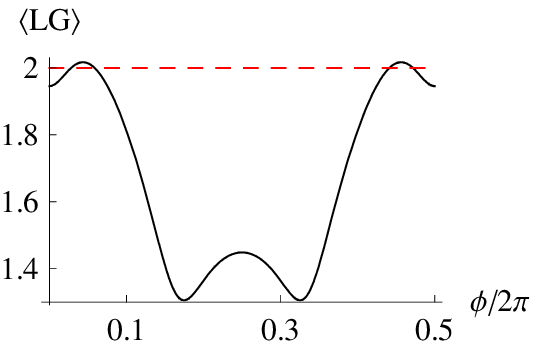,width=0.48\linewidth}
\label{fig:lgphi}}
\hfill
\subfigure[]{\epsfig{file=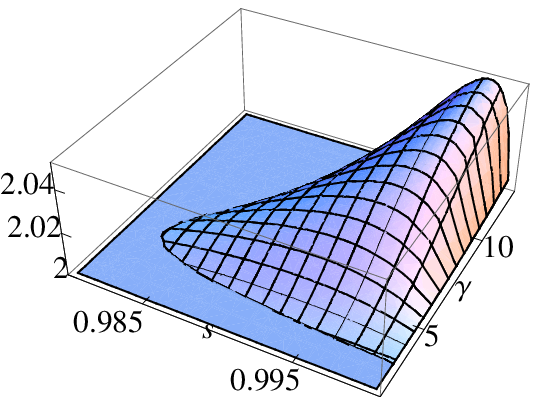,width=0.48\linewidth}
\label{fig:sgamma}}
\caption{\subref{fig:lgphi} A plot of the LG expectation value as a function of the time-interval $\phi$ between measurements (in units of oscillator period), for the step parameter $s=0.99$ and the squeezing parameter $\gamma\approx7.6$ that gives maximal violation for this $s$. The dashed red line corresponds the the bound $\expc{LG}=2$ imposed on realistic theories. \subref{fig:sgamma} A plot of $\max_\phi\expc{LG}$ as a function of $s$ and $\gamma$ (only values greater than $2$ are shown). The LGI violation is stronger for $s$ close 1 and $\gamma$ large. }
\end{figure}

\begin{figure}[hrb]
\subfigure[]{\epsfig{file=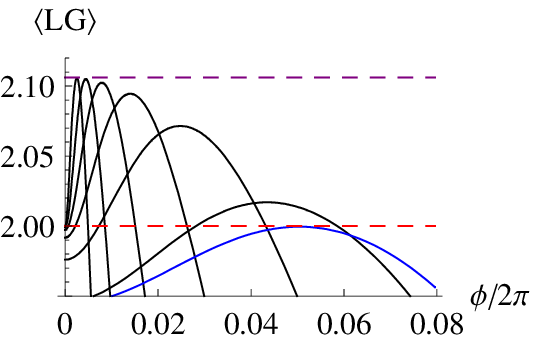,width=0.48\linewidth}
\label{fig:lgphi-sc}
}
\hfill
\subfigure[]{\epsfig{file=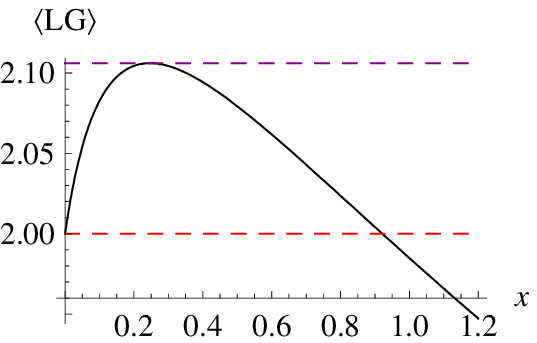,width=0.48\linewidth}
\label{fig:scaling}}
\caption{\subref{fig:lgphi-sc} A plot similar to Fig.\ \ref{fig:lgphi} with data shown for several values of $s$ between $s_{\text{cr}}\approx0.983$ (the minimal $s$ allowing LGI violation) and $s=1-10^{-7}$, and $\gamma$ that yields the maximal LGI violation for each $s$. $s$ is increasing in order from bottom to top.
The global maximum of $\expc{LG}$, $\approx2.106$, is shown in the upper dashed line in purple. 
\subref{fig:scaling}  The limit reached by $\expc{LG}$,  as  $s\to1$ $\gamma\to\infty$, $\phi\to0$, as a function of the scaling variable $x=\phi^2/(2\gamma^3(1-s)^2)$. The maximum is obtained for $x\approx0.24$}
\end{figure}

For this reason, we introduce a step parameter $-1\le s\le1$ such that when $s=-1$  $S$ is the singular (squeezed) parity observable, while $S$ becomes smooth when $1-s$ is small  (\cite{wodk} introduced similarly parametrized observables).  Our first main result is to show that the LGI is violated for a wide range of values $s$ and the squeezing parameter $\gamma$ (defined below), shown in Fig.\ \ref{fig:sgamma}. We next focus on the smooth limit $1-s\ll1$ and show that  the LGI violation persists in this domain, and is in fact {maximized} when $s\to1^-$, thus demonstrating breakdown of smooth observable MR in the harmonic oscillator. The maximal violation $\expc{LG}_\text{max}\approx2.106$ occurs in a singular limit where $s\to1$, $\gamma\to\infty$ and the time interval between the consecutive measurements tends to 0, where $\expc{LG}$ becomes a scaling function of a single parameter, shown in Fig.\ \ref{fig:scaling}. In contrast, if the smooth limit $1-s\to0$ is taken with fixed $\gamma$, regular semiclassical physics 
holds, showing a small violation of LGI for any $\gamma$ larger than a threshold value, implying breakdown of dynamic realism but not of MR. Lastly, by demonstrating the equivalence of our LGI test to a Bell-like measurement of a pair of entangled photons, we propose a quantum optical setup as an indirect method of observing LGI violations. Our scheme can be generalized to show that local realism in a composite system implies DR in a subsystem.

\paragraph{Violations of the Leggett-Garg inequality}
A harmonic oscillator offers a natural hidden trajectory description in terms of its classical trajectories. Choosing units such that the Hamiltonian $H=\omega\frac{q^2+p^2}{2}$, where $q$ and $p$ are the position and momentum of the oscillator respectively, and $\omega$ its frequency, the trajectories are $q_t=q\cos(\omega t)+p\sin(\omega t)$, $p_t=-q\sin\omega t+p\cos\omega t$. If we also prepare the oscillator in a state with a nonnegative Wigner function $W(q,p)$, such as a coherent state, the trajectories can be used to define a time-dependent phase-space probability density function $\mathsf{P}_t(q,p)=W(q_{-t},p_{-t})$ \cite{littlejohn}.

We therefore choose to test DR in a harmonic oscillator prepared in its ground state $\ket{g}$ by performing consecutive measurements of the operator $S=s^n$, where $n=\frac{A}{\hbar}-\frac12$, $A={\frac{1}{2}(\gamma^{-1}q^2+\gamma p^2)}$; $S$ depends on  the step parameter $|s|\le1$  and the squeezing parameter $\gamma>0$, and obeys $|S|\le1$.  $A$ is the action, and $n$ is the nonnegative integer-valued excitation number, of a squeezed harmonic oscillator with eigenstates $\ket{n}$. $s$ can also be viewed as a smoothness parameter: the relative change in $S$ between adjacent eigenstates is $1-s$, so that for $s\to1^-$ $S$ becomes smooth, and its measurement can be achieved with a classical apparatus that does not resolve individual states. In the opposite limit $s=-1$ $S$ becomes the singular parity operator that changes between $1$ and $-1$ for consecutive eigenstates. This property is also reflected in the Weyl representation of $S$,
$ \mathcal{S}(q,p)={\textstyle\frac{2}{1+s}} e^{\frac{(s-1)}{\gamma (s+1)  } \left(q^2+ \gamma^2p^2\right)}$.
$\mathcal{S}(q,p)$ is positive for any $s>-1$, but its maximum $\mathcal{S}_{\max}=\frac{2}{1+s}$ increases as $s$ decreases and diverges as $s\to-1$. In particular, since $\mathcal{S}_{\max}>\|S||=1$, $S$ is \emph{improper} \cite{technion} for any $s<1$, allowing for violations of realism.

The correlation function of two $S$ measurements
\begin{equation}
C(t)=\sum_{n'n}s^{n'+n}|\bra{n}e^{\frac{1}{i\hbar}Ht}\ket{n'}|^2|\braket{n'}{g}|^2
\end{equation}
depends only on the time difference $t$ because the initial state is an eigenstate of $H$. $C$ can be expressed as a contour integral over the unit circle
\begin{equation}\label{eq:cc}
C(t)=\oint\frac{dz}{2\pi iz}\bra{g}\rho(z)\ket{g}\Tr\rho(s)e^{\frac{Ht}{i\hbar}}\rho(s/z)e^{-\frac{Ht}{i\hbar}}\ ,
\end{equation}
where $\rho(x)=\sum_nx^n\ket{n}\bra{n}$ is a squeezed thermal state. The trace operations can be carried out for example using phase space integrations, and the result of the contour integration is
\begin{equation}\label{eq:c}
C(s,\gamma,\omega t)=\frac{c}{\sqrt{a-b_-}\sqrt{a+b_+}}K(m),
\end{equation}
where $K(m)=\int_0^1[(1-t^2)(1-mt^2)]^{-1/2}dt$ is an elliptic integral of the first kind,
\begin{align}
a&=\frac{\gamma+1}{(\gamma-1)}(4\gamma^2+(1-s^2)(\gamma ^2-1)^2 \sin^2(\omega t))\\
b_{\pm}&=4s^2\gamma^2\pm s(s^2-1)(\gamma^2-1)\sin(\omega t) \nonumber\\&
\qquad\times\sqrt{4\gamma ^2+(\gamma ^2-1)^2 \sin^2(\omega t)}\label{eq:b}\\
c&=\frac{8\gamma^{3/2}\sqrt{4\gamma ^2+(1-s^2)(\gamma ^2-1)^2 \sin^2(\omega t)}}{\pi  (\gamma-1 ) },
\end{align}
and $m=\frac{2 \left(b_--b_+\right) a}{\left(b_--a\right) \left(b_++a\right)}$. If we choose the three time intervals in the LG experiment to be of the same length then the LG expectation value is $\expc{LG}=3C(s,\gamma,\phi)-C(s,\gamma,3\phi)$, where $\phi=\omega t$.

$\expc{LG}$ is invariant under each of the transformations $s\rightarrow-s$, $\gamma\rightarrow\ 1/\gamma$ and  $\phi\rightarrow-\phi$, so our results are presented for positive values of $s$ and $\phi$ and for values of $\gamma$ greater than $1$. A typical graph of $\expc{LG}$ as a function of $\phi$ for fixed values of $s$ and $\gamma$ that allow LGI violation is shown in Fig. \ref{fig:lgphi}. The violation occurs in a small window of times that are significantly smaller than the period.
The region in the parameter plane that allows LGI violation is shown in Fig. \ref{fig:sgamma}. It is bounded by the rectangle 
$s>s_{\text{cr}}\approx0.98$ and $\gamma>\gamma_{\text{cr}}\approx3.2$.

The qualitative features of $\expc{LG}$ are as follows (i)  Fixing one of the parameters $s,\gamma$ at a value that allows LGI violation, the maximal violation occurs for a finite value of the other parameter and $\phi$.  (ii) $\max_\phi\expc{LG}$ increases as $s\to1$ and $\gamma\to\infty$, and $\phi\to0$, approaching the global maximum $\approx2.106$ (see Figs.\ \ref{fig:sgamma}, \ref{fig:scaling}). (iii) If one of the parameters is fixed, and another is taken to its extreme limit, $s\to1$ or $\gamma\to\infty$, the LGI holds.

\paragraph{Breakdown of macrorealism and the semiclassical limit}
If the trace in Eq.\ (\ref{eq:cc}) is evaluated as a phase-space integral, the maximization of $\expc{LG}$ can be understood as a trade-off between the overlap in the supports of the phase-space representations of the observable $S$ and its Heisenberg-picture time evolution $S_t$, that is maximized for the trivial values $s=1$, $\phi=0$, or $\gamma=1$, and the noncommutativity of $H$ and $S$, that is maximized for large $\gamma$. It follows that in order to maximize the LGI violation $s$ should approach 1 to optimize the phase-space overlap, and at the same time $\gamma$ should tend to infinity, and therefore also $\phi\to0$ to keep the phase-space overlap of the second measurement large.

This qualitative observation is in accord with the explicit functional behavior of $C$ as given in Eq.\ (\ref{eq:c}). Indeed the maximal LGI violation of $\expc{LG}$ is obtained when $s$ is close 1, but $s$ must approach $1$ in a singular limit where $\gamma\to\infty$, $\phi\to0$ according to $\gamma\sim(1-s)^{-1/2}$ and $\phi\sim(1-s)^{1/4}$. In this limit $\expc{LG}$ approaches a scaling form $\Lambda(x)=\frac2\pi\bigl(3K(-x)-K(-9x)\bigr)$ for the scaling variable $x=\phi^2/(2\gamma^3(1-s)^2)$. $\Lambda(x)$, shown in Fig.\ \ref{fig:scaling}, has a single maximum $\approx2.106$ (obtained at $x\approx0.24$), that is the global maximum of $\expc{LG}$.

It is remarkable that the maximal LGI violation, an unequivocal manifestation of quantum mechanics, is obtained in the limit when $S$ is smooth, and its measurement is well-modelled classically. In order to understand this surprising result it is instructive to compare it to the case where $s\to1^-$ and $\hbar\to 0$ together in such a way that $-\frac{\hbar}{\log s}$ tends to a finite limit $A_0$ with fixed $\gamma$ and $\phi$. In this case the Weyl representation of $S_t$, $e^{-\frac{A(q_t,p_t)}{A_0}+\frac{\hbar}{2A_0}}+O(\hbar^2)$ is smooth and regular semiclassical analysis is applicable so that LG violation is possible, but it can only be small, of $O(\hbar)$. Indeed, an explicit calculation gives
 $\expc{LG}=2+\bigl(f(\gamma)+g(\gamma)(3 \cos(2\phi)-\cos(6 \phi))\bigr)\frac{\hbar}{A_0}+O(\hbar)^2$, where $f(\gamma)=-\frac{\left(\gamma ^2+1\right) (\gamma-1)^4+4 \gamma \left(\gamma ^2-1\right)^2}{8 \gamma ^3}$, and $g(\gamma)=\frac{\left(\gamma ^2-1\right)^2 \left(1+\gamma ^2\right)}{16 \gamma ^3}$ . For $\phi=\pi/8$, and  $\gamma>\gamma_\text{cr}$
 $\expc{LG}$
 is larger than 2, but the LGI violation is small, and does not imply breakdown of MR.
 
This argument \emph{fails} in the scaling limit: Although $S$ changes slowly between adjacent states when $1-s\ll1$, its phase-space representation is a rapidly varying function when $\gamma\to\infty$, and naive semiclassical arguments fail. Put another way, although $S$ and $H$ are both smooth observables, they are mutually singular for large $\gamma$, so that a combination of a measurement of $S$ and an evolution with $H$ can break MR. 


\paragraph{Implementation with entangled photons}
Quantum optics offers a natural ground for implementing the LGI test proposed here. However, the standard method of counting photons is destructive, and prevents the second measurement in the LG setup from being carried out. We therefore propose to mimic the LG scheme using an entangled photon pair and a pair of number-resolving photon counters such that the results effectively measure the LG expectation value.

\begin{figure}[htb]
\subfigure{\epsfig{file=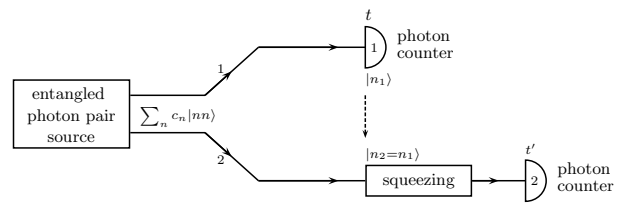,width=1\linewidth}}
\caption{\label{fig:photons}A schematic proposal for testing the LGI indirectly with a pair of entangled photons. The photon pair emitted from the source is in an entangled state $\sum_n c_n\ket{nn}$ with fully correlated photon number, where $c_n$ is the $n$-photon amplitude in the squeezed vacuum state. The number-resolving photon counter in the top arm effectively performs the first measurement in the LG scheme on the photon state in the bottom arm, which then undergoes the second measurement in the bottom photon counter. Different correlation functions can be measured by tuning the squeezing strength in the second mode.}
\end{figure}

Our proposed scheme is summarized in Fig.\ \ref{fig:photons}. We first reverse the roles of $H$ and $S$, such that $H$ is the squeezed Hamiltonian, and $S=s^{n_\text{photons}}$. The $H$ ground state is then a squeezed vacuum state $\ket{g}=\sum_nc_{2n}\ket{2n}$, where $c_{2n}=\frac{\sqrt{2} \gamma ^{1/4} }{\sqrt{(1+\gamma)}}\frac{(2n-1)!!}{\sqrt{(2 n)!}}\bigl(\frac{\gamma -1}{\gamma +1}\bigr)^n$. Our goal is to obtain in the wake of a measurement of $n$ photons a collapsed state $\ket{n}$. This is not practical for a single photon, but the same effect can be realized if the initial state is a \emph{two-photon} entangled state $\sum_{n=\text{even}}c_n\ket{nn}$, $\ket{nn}$ being a state with $n$ photons in both modes. A measurement of $n_1$ photons (with probability $|c_{n_1}|^2$) in the first mode collapses the second to the Fock state $\ket{n_1}$. Next, time-evolution by the squeezed Hamiltonian is applied to the state in the second arm, and finally the number of photons in the second mode $n_2$ is also measured. Since the photon numbers are perfectly correlated, the value of $s^{n_1+n_2}$, averaged over a large number of repetitions constitutes a measurement of one of the two correlation functions needed for the LG expectation value, and the second correlation function is obtained by changing the squeezing interaction.

In principle, MR would be tested for strongly squeezed initial states that has many significant photon state components, but 
current number-resolving photon counters can detect only a few photons. A more fundamental limitation is placed by decoherence as discussed below.

An experiment that realizes our proposed scheme is only an indirect demonstration of the violation of DR, since it measures violations of the LGI in a hypothetical system whose dynamics should follow the one studied in the experiments according to the
rules of quantum mechanics. 
On the other hand, the casting of the LGI into inequalities bounding expectation values in an entangled pair has further consequences of a more fundamental nature. It is recognized that in the setting of the Bell inequalities and their generalizations, the measurements by the two parties may be performed at an arbitrary time delay. Furthermore, the combination of the squeezing action and the second photon counting is equivalent to a measurement of a different observable, unitarily equivalent to $S$, so that the squeezing is analogous to the rotation of the apparatus in the Bell scheme. The upshot is that the violation of the LGI is tantamount to a violation of the CHSH inequality in an associated system such as the one shown in Fig.\ \ref{fig:photons}. Hence DR in a subsystem is implied by local realism in the composite system. 

The implication of DR by local realism has been demonstrated in the specific context of an entangled pair of  harmonic oscillators, but since the demonstration did not rely on specific properties of harmonic oscillators, it holds for any quantum system that can be entangled with another system and prepared in a diagonal state like the one shown in  Fig.\ \ref{fig:photons}. On the other hand, we conjecture that the converse statement is false, since the Hilbert space dimension of  entangled states is greater than that of the subsystem.

Recently there have been several studies of the breakdown of phase-space local realism in two-photon entangled states \cite{wodk,bw,chen,wilson,teich}. In these works the breakdown of local realism was demonstrated using classically singular observables related to the number parity operator. In contrast, we demonstrate breakdown of {MR} and local realism by measurements of classically smooth observables.

\paragraph{Conclusions}
Our study has shed some light on the persistent issue of the emergence of classical realism in closed quantum systems, specifically the breakdown of MR by the measurement process. We have shown that the principles of classical mechanics cannot describe even the measurement of macroscopic observables with a smooth classical limit.

The results further restrict the validity of MR in closed systems beyond the analysis of \cite{kofler1,kofler2}. These works focused on the coarse-grained measurement of singular observables, and showed that MR can be broken for singular Hamiltonians. Here the macroscopic nature of the measurement process is realized with smooth observables, and MR is broken for a Hamiltonian that is itself smooth. This is possible because the Hamiltonian and the observable to be measured are mutually singular, but this is not in principle an obstruction to a realization of the measurement and dynamics with a classical apparatus. Still, a macroscopic object is necessarily an open system, and if this is
taken into account,
breakdown of MR can be avoided if decoherence is strong enough. If this is indeed the case, our results imply a minimum rate of decoherence. Namely, since violation of MR occurs on a time scale inversely proportional to $\sqrt{n_\text{init}}$, where $n_\text{init}$ is the number of $S$ eigenstates in the initial state, the time scale for decoherence must grow faster than $\sqrt{n_\text{init}}$, when $n_\text{init}\gg1$.

Since the present 
violation of the LGI stems from the counterfactual assumption of the possibility of noninvasive measurements, the question naturally arises of whether weak MR, where this assumption is relaxed to allow invasive measurements that admit a hidden-variable underpinning, is valid in the harmonic oscillator. It has been recently shown \cite{zela} that weak MR also has observable consequences.
 
A further conclusion from this work is that there is a logical relation between the postulates of DR and those of local realism, that facilitates an indirect experimental test of  MR with an entangled photon pair. It is likely connected to a general relation between temporal and spatial nonclassicality that has been derived in \cite{reznik} by focusing  on the unitary evolutions rather than on the quantum states.

We thank H. Eisenberg and B. Reznik for helpful discussions. This work was supported by the GIF grant 980-184.14/2007.

\end{document}